\documentclass[twocolumn,prl,showpacs]{revtex4}
\usepackage{psfrag}
\usepackage{graphicx}
\usepackage{amsmath}
\usepackage{amssymb}

\begin{document}
\title{Partly Occupied Wannier Functions}
\author{K. S. Thygesen}
\affiliation{Center for Atomic-scale Materials Physics, \\
  Department of Physics, Technical University of Denmark, DK - 2800
  Kgs. Lyngby, Denmark}
\author{L. B. Hansen}
\affiliation{Center for Atomic-scale Materials Physics, \\
  Department of Physics, Technical University of Denmark, DK - 2800
  Kgs. Lyngby, Denmark}
\author{K. W. Jacobsen}
\affiliation{Center for Atomic-scale Materials Physics, \\
  Department of Physics, Technical University of Denmark, DK - 2800
  Kgs. Lyngby, Denmark}
\date{\today}

\begin{abstract}
  We introduce a scheme for constructing partly occupied, maximally
  localized Wannier functions (WFs) for both molecular and periodic
  systems.  Compared to the traditional occupied WFs the partly
  occupied WFs posses improved symmetry and localization properties
  achieved through a bonding-antibonding closing procedure.  We
  demonstrate the equivalence between bonding-antibonding closure and
  the minimization of the average spread of the WFs in the case of a
  benzene molecule and a linear chain of Pt atoms. The general
  applicability of the method is demonstrated through the calculation
  of WFs for a metallic system with an impurity: a Pt wire with a
  hydrogen molecular bridge.
\end{abstract}
\maketitle

The idea of representing the electronic ground state of a solid-state
system in terms of localized orbitals dates back to
Wannier~\cite{wannier37} who introduced a canonical transformation
between the Bloch states of an insulating crystal and a set of
localized functions. These localized Wannier functions (WFs) provide a
formal justification of the widely used
tight-binding~\cite{ashcroft_mermin} and Hubbard models~\cite{mahan}.
A scheme for evaluation of maximally localized WFs for composite bands
has been developed by Marzari and Vanderbilt~\cite{marzari97}, and the
WFs have turned out to be useful in a number of contexts including
electron transport calculations~\cite{calzolari04}, linear-scaling
electronic structure methods~\cite{goedecker99}, and in the theory of
electric polarization~\cite{king-smith93,resta94}.

In the context of molecular systems the analogous of Wannier functions
for finite systems has been studied under the name "localized
molecular orbitals"~\cite{boys60, foster60, edmiston63, pipek89,
  berghold00, silvestrelli98}.  These are traditionally defined by an
appropriate unitary transformation of the occupied single-particle
eigenstates and represent a strong analysis tool as they provide
insight into chemical properties such as bond types, coordination, and
electron lone pairs.  In the following we shall for simplicity use the
term WF to cover also localized molecular orbitals.

\begin{figure}[!b]
\includegraphics[width=0.88\linewidth]{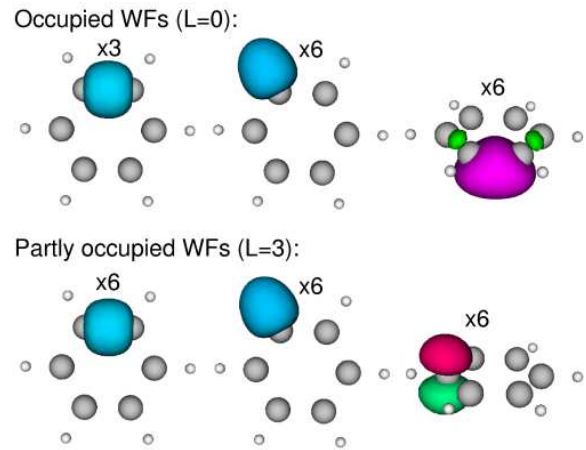}
\caption[cap.Pt6]{\label{fig.benzene} Occupied and partly occupied Wannier
  functions for benzene $\text{C}_6\text{H}_6$. The traditional
  (occupied) WFs are seen to mix the $\sigma$ and $\pi$ orbitals.
  Inclusion of the three anti-bonding $\pi$ orbitals leads to the partly
  occupied WFs which are more localized and separate the orbitals
  with different mirror symmetry.}
\end{figure}

In this letter we demonstrate that in constructing localized WFs it is
in some cases advantageous to consider not only the space of occupied
states but to include also selected unoccupied states through a
bonding-antibonding closing procedure.  The changes achieved by the
bonding-antibonding closing can be viewed directly in
Fig.~\ref{fig.benzene} where calculated WFs with and without the
inclusion of additional unoccupied states are shown in the case of a
benzene molecule.  In the upper panel only the occupied states are
used and some of the resulting WFs clearly mix $\sigma$- and
$\pi$-character.  By including three selected unoccupied states the
resulting partially occupied WFs become more localized and the
$\sigma$- and $\pi$-systems are now separated. In addition the
rotational symmetry of the molecule is more clearly reflected by the
partly occupied WFs.

The problem solved by the bonding-antibonding closing is the
following: if we consider two well-localized atomic orbitals on
neighboring atoms in a molecule and allow the two states to hybridize,
a bonding and an antibonding combination will result -- combinations
which may be less localized than the individual atomic orbitals.  To
regain the localized atomic orbitals from the molecular orbitals we
need both the bonding and the antibonding combination independent of
their occupation. In some cases the antibonding state may have
hybridized further with other states and the state which ``matches''
the bonding state will be a linear combination of eigenstates. To
apply the bonding-antibonding closing in practice a systematic scheme
to identify the relevant unoccupied states is clearly required. We
show that this can be achieved by optimizing the localization of the
resulting WFs.

The lack of localization of WFs obtained if only occupied states are
considered is well-known in the case of extended metallic
systems~\cite{souza01, iannuzzi02}. A proper localization of the WFs
requires in this case completion of the partly occupied valence
bands (the analogue of bonding-antibonding closing for a periodic
system) which may be difficult due to crossings and mixing of the
valence bands with unwanted higher-lying bands. This problem was
addressed by Souza, Marzari, and Vanderbilt~\cite{souza01} who
constructed a disentangling procedure which makes it possible to
follow a given band by minimizing the change in character of the Bloch
states across the Brillouin zone. In this way the relevant unoccupied
states could be identified. The approach we suggest here makes no
reference to the wave vector, it is simple to apply and
valid for both isolated and periodic systems.

We note that different techniques for constructing
localized orbitals have been developed in connection with linear
scaling electronic structure methods~\cite{goedecker99}. 
In that context the orbitals are typically constructed during an
electronic structure calculation and are required to
vanish outside predefined localization regions. In contrast, the
scheme we propose is an exact transformation of already determined
eigenstates to obtain maximal localization.

Our starting point is a set of single-particle eigenstates
$\{\psi_n\}$ obtained for example from a Density Functional Theory
(DFT)~\cite{dft} calculation. We denote the total number of
eigenstates included in the DFT calculation by $N$ and the number of occupied eigenstates by $M$.
Following Marzari and Vanderbilt~\cite{marzari97} we measure the
spread of the WFs $\{w_n\}$ by the sum of second moments
\begin{equation}\label{eq.spreadfct1}
S=\sum_{n} (\langle w_n | r^2 | w_n \rangle - \langle w_n |\mathbf{r}
| w_n \rangle ^2).
\end{equation}
The WFs are expanded in terms of the $M$ occupied eigenstates and $L$
extra degrees of freedom (EDF) $\{\phi_l\}$ from the
$(N-M)$-dimensional space of unoccupied eigenstates. This will lead to
a total of $M+L$ WFs and we thus require $N\geq M+L$.
The expansion takes the form
\begin{equation}\label{eq.expansion1}
w_n=\sum_{m=1}^{M}U_{mn}\psi_m+\sum_{l=1}^{L}U_{M+l,n}\phi_l,
\end{equation}
where the unoccupied orbitals are written as
\begin{equation}\label{eq.expansion2}
\phi_l=\sum_{m=1}^{N-M}c_{ml}\psi_{M+m}.
\end{equation}
The columns of the matrix ${c}$ are taken to be orthonormal and
represent the coordinates of the EDF with respect to the unoccupied
eigenstates. The matrix $U$ is unitary and represents a rotation in
the space of the functions
$\{\psi_1,\ldots,\psi_M,\phi_1,\ldots,\phi_L\}$.
Through the expansions (\ref{eq.expansion1}) and
(\ref{eq.expansion2}), $S$ becomes a function of $U_{ij}$ and
$c_{ij}$, which must be minimized under the constraint that $U$
is unitary and the columns of $c$ are orthonormal. 

To keep technicalities at a minimum the formulation given above has
been specialized to systems which are either isolated or described in
a periodically repeated supercell in the $\Gamma$-point approximation.
A detailed account of the method including the $\mathbf k$-point
formulation will be given elsewhere~\cite{2bpublished}. Here we focus
on the case of a cubic supercell of length $a$ in the $\Gamma$-point
approximation. Extension of the theory to supercells of general
symmetry is straightforward along the lines of
Refs.~\onlinecite{silvestrelli99} and \onlinecite{berghold00}.

In the limit of a large supercell, minimizing $S$ is equivalent to
maximizing the functional~\cite{berghold00,resta99}
\begin{equation}\label{eq.spreadfct2}
\Omega =\sum_{n}  (|X_{nn}|^2+|Y_{nn}|^2+|Z_{nn}|^2),
\end{equation}
where the matrix ${X}$ is defined as $X_{nm}=\langle
w_n|e^{-i(2\pi/a)x}|w_m \rangle$, with similar definitions for
$Y_{nm}$ and $Z_{nm}$.  The unitary matrix at iteration $n$ is written
as $U^{(n)}=U^{(n-1)}\text{exp}(-A)$, where $A$ is an anti-hermitian
matrix. By straightforward differentiation we obtain expressions for
the gradients $\partial \Omega/\partial A_{ij}$ and $\partial
\Omega/\partial c^*_{ij}$ giving the steepest uphill direction of
$\Omega$ upon variations in $A$ and $c$. The orthonormality constraint
on $c$ is invoked through a set of Lagrange multipliers.  Any
gradient-based optimization scheme can now be applied to maximize
$\Omega$. 
 
Given $N$ and $L$, the algorithm introduced above produces the $M+L$
most localized WFs. We shall assume that $N$ has been fixed at a value large
enough to include all anti-bonding states relevant for
the localization. The problem is then to determine a
good value for $L$. 
It seems as a natural strategy to choose an $L$ that gives a high localization
\emph{per} orbital. To quantify this statement we define the average
localization per orbital as
\begin{equation}\label{eq.averagespread}
\langle \Omega \rangle=\frac{\Omega}{M+L},
\end{equation}
where $M+L$ is the total number of WFs. From the definition of
$\Omega$ it is clear that $0\leq \langle \Omega \rangle \leq 3$.
Fixing $L$ on the basis of $\langle \Omega \rangle$ represents a
quite general criterion which can be applied to any system. However,
the localization procedure must be
carried out for several $L$ which might be a
tedious task depending on the size of the system. 
Formally the global maximum of $\langle \Omega \rangle$ is attained
in the limit where both $N$ and $L$ tend to infinity in which case
completely localized delta functions can be realized. 
However, we have found that for practical values of $N$ where
very high energy states are not considered, $\langle
\Omega \rangle$ goes through a local maximum as $L$ is increased beyond
0. For many systems it is possible to determine
a value for $L$ based on symmetry arguments, chemical intuition, or a
closed band condition.  In particular the latter strategy is natural
for crystalline systems where the closed band condition is satisfied
by choosing $L$ equal to the number of unoccupied states belonging to
all partly occupied bands. As we shall see in the following examples
the two criteria for determining $L$ lead to similar results.

All the following calculations are based on the Kohn-Sham eigenstates
as defined within DFT~\cite{dft}. We use a plane wave based
pseudopotential code~\cite{dacapo} with an energy cutoff of 25 Ry for
the plane wave expansion and describe the ion cores with ultrasoft
pseudopotentials~\cite{vanderbilt90}.

We have constructed the maximally localized WFs of an isolated benzene
molecule in a cubic supercell of length $a=16$ \AA. If we consider
only the occupied eigenstates of benzene, corresponding to $L=0$ in
the method, we find the 15 occupied WFs shown in the upper panel of
Fig.~\ref{fig.benzene}. Nine of these are $\sigma$-bonds centered at
each C-H and every second C-C bond, and the last six orbitals are
double bonds with a mixed $\sigma$- and $\pi$-character centered at
the three remaining C-C bonds.  This is the classical picture of the
chemical bonds in benzene. In this picture the resulting orbitals
apparently break the six-fold rotational symmetry of the molecule,
i.e. the bonding orbitals between different C-C pairs differ.
Moreover, the planar structure of the molecule suggests a separation
of $\sigma$- and $\pi$-orbitals contrary to C-C double bonds. The
occupied subspace spanned by all the WFs of course posseses the full
symmetry.

\begin{figure}
  \includegraphics[width=0.75\linewidth,angle=270]{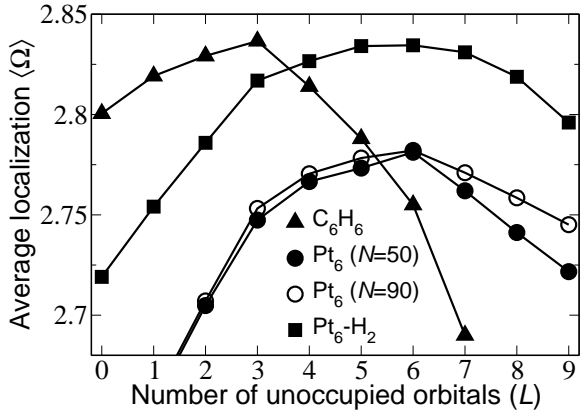}
  \caption[cap.Pt6]{\label{fig.averagespread} 
    Average localization per Wannier function as a function of the
    number of included unoccupied orbitals for the three systems
    studied. For the benzene molecule the maximum appears at $L=3$
    corresponding to the inclusion of the three antibonding $\pi$
    orbitals. For the Pt chain the maximum at $L=6$ corresponds to
    completion of the valence bands. (The values of $\langle \Omega
    \rangle$ do not compare from one system to another due to
    dependence on the size of the supercell.)}
\end{figure}

\begin{figure}
\includegraphics[width=0.8\linewidth,angle=270]{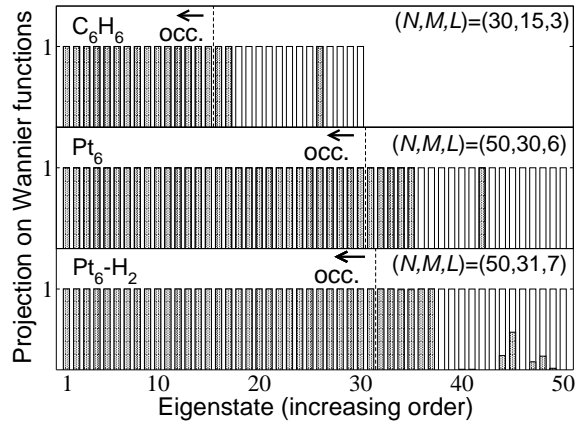}
\caption[cap.project]{\label{fig.project} 
  The norm of the projection of eigenstates onto the subspace spanned
  by the partly occupied WFs for the three systems studied. The
  eigenstates are listed in increasing order according to their
  energy. States below the dashed line are occupied. It is seen that
  in the case of the benzene molecule and the Pt chain the algorithm
  selects particular unoccupied states necessary for the
  bonding-antibonding closing. In the case of the hydrogen molecule in
  the Pt chain one of the WFs has weight on an unoccupied resonance
  distributed over several eigenstates.}
\end{figure}

The lack of symmetry in the chemical picture of benzene is a result of
ignoring the three anti-bonding $\pi$-orbitals of the C ring in the
construction of the WFs. In Fig.~\ref{fig.averagespread} (triangles)
we plot the average localization of the WFs for various $L$ values.
The maximum occurs for $L=3$. To analyze the composition of the EDF we
have projected the molecular eigenstates onto the WFs generated with
$L=3$, see upper panel of Fig.~\ref{fig.project}. By inspection we
have verified that the EDF selected by the algorithm are exactly the
three anti-bonding $\pi$-orbitals of the C ring. Maximizing $\langle
\Omega \rangle$ is therefore equivalent to performing a
bonding-antibonding closure.  The 18 partly occupied WFs obtained with
$L=3$ consist of $\sigma$-bonds at all C-H and C-C pairs and a
$p$-orbital perpendicular to the molecular plane centered at each C
atom (lower panel of Fig.~\ref{fig.benzene}).  Consequently, the
inclusion of 3 EDF restores the rotational symmetry and separates
$\sigma$- and $\pi$-orbitals.

\begin{figure}
\includegraphics[width=1.0\linewidth]{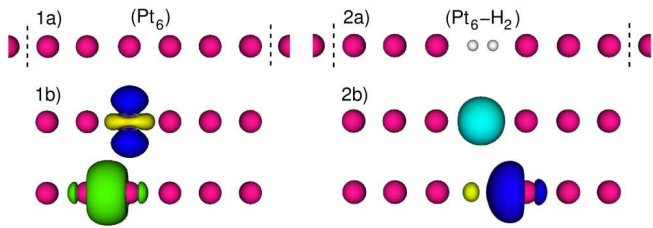}
\caption[cap.Pt6_states]{\label{fig.Pt6} Partly occupied Wannier
  functions for a periodically repeated, linear string of 6 Pt atoms
  without and with an inserted hydrogen bridge. The WFs
  have been generated with $L=6$ ($\text{Pt}_6$) and $L=7$
  ($\text{Pt}_6\text{-H}_2$). The WFs shown for the
  $\text{Pt}_6\text{-H}_2$ system correspond to $\text{H}_2$ and H-Pt
  $\sigma$-bonds.}
\end{figure}

We have performed a similar analysis for a linear chain of six Pt
atoms with periodic boundary conditions ($\text{Pt}_6$), see
Fig.~\ref{fig.Pt6}(1a). Such monatomic chains have been produced and
studied in STM experiments~\cite{ohnishi98} and by the mechanically
controlled break junction technique~\cite{smit_origin01}.

The valence bands of an infinite Pt wire consist of five $d$- and one
$s$-band. For each band only six Bloch states comply with the boundary
conditions in $\text{Pt}_6$ and consequently it takes a total of 36
states to close the bands. Since there are 30 occupied states, the
closed band condition can thus be fulfilled with $L=6$. From
Fig.~\ref{fig.averagespread} (circles) we see that the closed band
condition coincides with the maximum of $\langle \Omega \rangle$. In
the middle panel of Fig.~\ref{fig.project} we show the projection of
eigenstates onto the Wannier functions corresponding to $L=6$. By
comparing with the band diagram of the Pt wire, we find a perfect
match between the EDF and the unoccupied Bloch states of the valence
bands. Again this shows that closing the bands is equivalent to
minimizing the average spread. For $L=6$ we find five atom-centered
$d$-orbitals, of which one is shown, and one
$\sigma$-orbital centered between each Pt-Pt pair. The transformation
to WFs allows for explicit extraction of tight-binding parameters. For example, the doubly
degenerate $\delta$-band of the Pt chain is characterized by the hopping
parameters (in eV) $t_1=1.08$, $t_2=0.13$, $t_3=0.03$.

When $N$ is increased from 50 to 90 the localization generally
improves due to the larger space available for the EDF, see
Fig.~\ref{fig.averagespread} (open circles). If we increase $N$ further
this trend would continue and eventually the maximum would
start shifting towards larger $L$ values. In the limit $N\to \infty$
$\langle \Omega \rangle$ would increase monotonically as a function of
$L$. This shift of the maximum, however,
occurs only when $N$ becomes very large, i.e. when high-energetic
and chemically irrelevant states enter the localization space.   

As a final example we analyze the effect on the WFs when a molecular
hydrogen bridge is inserted in the $\text{Pt}_6$ system, see
Fig.~\ref{fig.Pt6}(2a). A molecular hydrogen bridge between Pt
contacts has recently been realized and studied
experimentally~\cite{smit01}. The average localization has a maximum
for $L=6$. The maximum is, however, not very pronounced and its
position depends on the Pt-H and H-H bond lengths. Consequently we
have studied the WFs for $L=5,6,7$. In all three cases we recover the
five $d$-orbitals per Pt found for $\text{Pt}_6$. The
$\sigma$-orbitals are slightly distorted and the one in the bridge
region is replaced by: an $\text{H}_2$ bonding orbital ($L=5$), two
$s$-orbitals located symmetrically on each H atom ($L=6$), an
$\text{H}_2$ bonding orbital and two Pt-H bonds with $\sigma$ symmetry
($L=7$). The latter situation is shown in Fig.~\ref{fig.Pt6}(2b). As
can be seen from the lower panel of Fig.~\ref{fig.project} this system
provides an example where one of the EDF is a superposition of
eigenstates.

To conclude we have proposed a method for finding maximally localized,
partly occupied WFs. The method ensures by construction that any
occupied eigenstate can be expanded in terms of the obtained WFs.  The
inclusion of additional unoccupied states in the definition of the WFs
improves the localization and simplifies the resulting chemical
pictures.  The method can easily be modified so that not the occupied
states but for example the states in an energy window around the Fermi
level are exactly represented. Extra high- and low-energy states can
then be included to construct effective model Hamiltonians for the
states around the Fermi level.

We acknowledge discussions with S{\o}ren L. Frederiksen and support
from the Danish Center for Scientific Computing through Grant No.
HDW-1101-05.


\bibliographystyle{apsrev}

\end{document}